\documentclass[aps,prl,reprint,10pt,superscriptaddress,longbibliography,nobibnotes]{revtex4-2}
\usepackage{tikz}
\usetikzlibrary{calc}
\usepackage{amsmath,amssymb,bm,booktabs,mathtools,microtype,longtable,needspace,xurl}
\usepackage[hidelinks]{hyperref}
\newcommand{\R}{\mathbb R}

\newcommand{\norm}[1]{\left\lVert #1\right\rVert}

\hypersetup{pdftitle={On the Seven-particle Clusters of Lennard-Jones and Thomson},pdfauthor={Li Shen}}
\begin{document}
\title{On the Seven-particle Clusters of Lennard-Jones and Thomson}
\author{Li Shen}
\affiliation{Department of Mechanical Engineering, Imperial College London, London SW7 2AZ, United Kingdom}
\begin{abstract}
In this Letter, we develop a graph-geometric framework and prove that the pentagonal bipyramid (PBP) is the unique global minimiser of both the seven-particle Lennard–Jones energy in \(\mathbb{R}^3\) and the seven-charge Thomson energy on \(S^2\), up to rigid symmetries and particle relabelling. For the Thomson problem, this strengthens a recent deterministic global-optimality certificate to numerical tolerance into an exact global-optimality and uniqueness theorem. A \textit{weak-interaction graph}, defined by thresholding pair geometry, organises configuration space without imposing symmetry or discarding any pair interaction. For Lennard-Jones, pair-energy counting, graph classification, subset bounds, and force balance reduce all possible minimisers to 29 bounded domains; exact geometric certificates exclude 28, localise the survivor near the PBP, and a uniform restoring-energy estimate proves uniqueness. The Thomson proof follows the same progression: graph reduction, geometric localisation, and rigidity, but finalises through whole-domain coercivity and a global localisation theorem. The same fivefold order therefore emerges in two very different physical settings: an unconstrained cluster with a preferred bond length and purely repelling charges confined to a sphere, showing a common geometric organising principle for seven particles that is not tied to either interaction.

\end{abstract}
\maketitle
\raggedbottom

How can a rotationally invariant pair interaction for $N$ identical particles uniquely select a structure with fivefold order? For $N>4$, geometry prevents all pair separations from reaching the Lennard-Jones (LJ) equilibrium distance simultaneously, so the ground state is a frustrated compromise among all $\binom{N}{2}$ interactions. The $D_{5h}$ PBP has long been the putative LJ$_7$ ground state, from early small-cluster studies to modern numerical searches \cite{HoarePal1971,wales}, but such searches cannot exclude the continuum of distorted and asymmetric configurations required for a proof of global optimality. An early deterministic scheme reported optimality certificates including $N=7$ \cite{MaranasFloudas1992}; its $N=7$ certificate was later shown to be invalid because of a nonphysical algorithmic parameter choice \cite{kuznetsov}. The strongest subsequent general-purpose spatial branch-and-bound certification reaches $N=6$ \cite{kuznetsov}, while stronger diameter bounds alone do not overcome this frontier \cite{lecomte2026}. The $N=7$ problem therefore requires a structural treatment of the competing low-energy configurations, rather than merely a tighter spatial search. The spherical Coulomb analogue has the same organising obstruction: its 21 pair distances must come from one common rank-three Gram matrix. We return to that problem after completing the LJ proof.

\emph{Lennard--Jones theorem.} The seven-particle ground state is the pentagonal bipyramid, unique up to Euclidean isometries and particle relabelling. We first construct an exact stationary PBP candidate $Y$. The proof then bounds all possible minimisers and uses geometric certificates to confine them to a neighbourhood of $Y$. A uniform local energy estimate shows that every configuration in this neighbourhood not congruent to $Y$ has higher energy, thereby establishing global optimality and uniqueness. We begin by stating the problem and the reference structure:

Working in reduced LJ units, so that the pair potential has minimum -1 and separation 1, we consider $X=(x_1,\ldots,x_N)$ to be a configuration of distinct points in $\mathbb{R}^3$. The $N$-particle energy is
\begin{equation}
E_N[X]=\sum_{i<j}V(r_{ij}),\quad
V(r)=\frac{1}{r^{12}}-\frac{2}{r^{6}},
\label{eq:energy}
\end{equation}
where $r_{ij}=\norm{x_i-x_j}$. The ground state is a configuration of $N$ particles with the lowest possible value of $E_N$. We prove that for $N=7$, a unique ground state exists up to Euclidean isometry and particle relabelling.

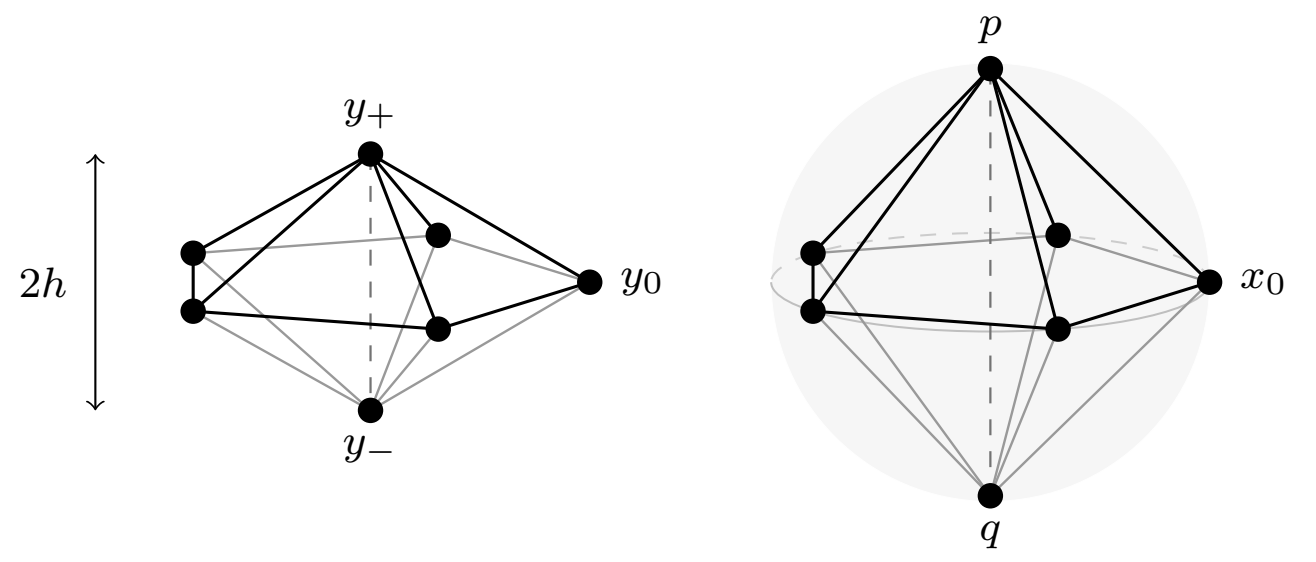
\begin{figure}[t]
\centering
\begin{tikzpicture}[
    font=\footnotesize,
    particle/.style={
        circle,
        fill=black,
        draw=black,
        inner sep=1.55pt
    },
    frontedge/.style={
        draw=black,
        line width=0.58pt
    },
    backedge/.style={
        draw=black!40,
        line width=0.45pt
    },
    axisline/.style={
        draw=black!55,
        line width=0.42pt,
        dashed
    },
    sphereguide/.style={
        draw=black!25,
        line width=0.38pt
    },
    sphereback/.style={
        draw=black!20,
        line width=0.38pt,
        dashed
    }
]

\def\R{1.45}
\def\viewang{77}

\pgfmathsetmacro{\cv}{cos(\viewang)}
\pgfmathsetmacro{\sv}{sin(\viewang)}

\begin{scope}[xshift=-2.05cm]

\pgfmathsetmacro{\H}{\R*(0.5112/0.8518)}

\foreach \i in {0,...,4}{
    \pgfmathsetmacro{\a}{72*\i}
    \coordinate (L\i) at
        ({\R*cos(\a)},{\R*\cv*sin(\a)});
}

\coordinate (LP) at (0,{ \H*\sv});
\coordinate (LM) at (0,{-\H*\sv});

\draw[backedge] (L0)--(L1)--(L2);

\foreach \i in {0,...,4}{
    \draw[backedge] (LM)--(L\i);
}

\draw[axisline] (LP)--(LM);

\draw[frontedge] (L2)--(L3)--(L4)--(L0);

\foreach \i in {0,...,4}{
    \draw[frontedge] (LP)--(L\i);
}

\foreach \i in {0,...,4}{
    \node[particle] at (L\i) {};
}
\node[particle] at (LP) {};
\node[particle] at (LM) {};

\node[above=1.5pt] at (LP) {$y_+$};
\node[below=1.5pt] at (LM) {$y_-$};
\node[right=2.5pt] at (L0) {$y_0$};

\draw[<->,thin]
    ($(LM)+(-1.82,0)$)
    -- node[left=2pt] {$2h$}
    ($(LP)+(-1.82,0)$);

\node[below=12pt] at (0,-\R) {(a) LJ$_7$};

\end{scope}

\begin{scope}[xshift=2.05cm]

\foreach \i in {0,...,4}{
    \pgfmathsetmacro{\a}{72*\i}
    \coordinate (T\i) at
        ({\R*cos(\a)},{\R*\cv*sin(\a)});
}

\coordinate (TP) at (0,{ \R*\sv});
\coordinate (TM) at (0,{-\R*\sv});

\fill[gray!7] (0,0) circle (\R);
\draw[gray!0,line width=0pt]
    (0,0) circle (\R);

\draw[sphereback]
    (\R,0)
    arc[
        start angle=0,
        end angle=180,
        x radius=\R,
        y radius={\R*\cv}
    ];

\draw[sphereguide]
    (-\R,0)
    arc[
        start angle=180,
        end angle=360,
        x radius=\R,
        y radius={\R*\cv}
    ];

\draw[backedge] (T0)--(T1)--(T2);

\foreach \i in {0,...,4}{
    \draw[backedge] (TM)--(T\i);
}

\draw[axisline] (TP)--(TM);

\draw[frontedge] (T2)--(T3)--(T4)--(T0);

\foreach \i in {0,...,4}{
    \draw[frontedge] (TP)--(T\i);
}

\foreach \i in {0,...,4}{
    \node[particle] at (T\i) {};
}
\node[particle] at (TP) {};
\node[particle] at (TM) {};

\node[above=1.5pt] at (TP) {$p$};
\node[below=1.5pt] at (TM) {$q$};
\node[right=2.5pt] at (T0) {$x_0$};

\node[below=12pt] at (0,-\R) {(b) Thomson $T_7$};

\end{scope}

\end{tikzpicture}

\caption{Pentagonal-bipyramid geometries for the seven-particle
Lennard--Jones and Thomson problems, shown under the same orthographic
projection. In (a), the Lennard--Jones candidate has
$h/\rho \simeq 0.600$ and is therefore flatter than the spherical
pentagonal bipyramid. In (b), the five ring particles lie on an
equatorial great circle of the unit sphere and the two axial particles
are antipodal poles. Pale lines indicate the rear part of the
configuration; all lines are geometric guides.}
\label{fig:PBP-both}
\end{figure}

This is the pentagonal bipyramid (PBP), denoted by $Y=(y_0,\ldots,y_4,y_+,y_-)$, where $y_k=\rho (\cos\theta_k,\sin \theta_k,0)$ and $y_\pm=(0,0,\pm h)$ for $k=0,\ldots,4$ and $\theta_k=2\pi k/5$ and $\rho,h>0$. Numerically, $E_7\approx-16.5054$. Minimising the PBP energy over its shape and scale gives a unique positive stationary configuration, from which we obtain $\rho\approx0.8518$ and $h\approx0.5112$ (SM Sec.~S1). For global exclusions below, we use the certified cutoff value (SM Sec.~S1) 
\begin{equation}
U=-16.505384168012215592988>E_7(Y).
\label{eq:lj-cutoff}
\end{equation}
Now, stationarity within the PBP family, i.e. $\partial_\rho E_Y=\partial_h E_Y=0$, implies stationarity in the full Cartesian configuration space. At $Y$, symmetry determines all 21 Cartesian components of $\nabla E_7$ from only two scalars. For each ring particle, $y_k$, the axial force component is eliminated by reflection in the equatorial plane, while the tangential component disappears because of reflection in the vertical plane through the particle. The force on each ring particle is therefore purely radial, and fivefold rotation makes its radial component the same at all five ring particles due to rotational symmetry. At either pole, fivefold rotational symmetry constrains the force to be on the symmetry axis, and equatorial reflection ensures the two poles' forces are equal and opposite. So the only possible nonzero force components are the common ring-radial and pole-axial components. Varying $\rho$ tests the former and varying $h$ tests the latter, and so the two PBP stationarity conditions imply that every particle has zero net force, and so $\nabla E_7(Y)=0$. Thus $Y$ is an exact stationary configuration. We now turn to the global argument, which applies to arbitrary seven-particle configurations. 

\emph{Step 1: bound all possible minimisers. }Define a configuration $X$ to be \textit{competitive} if $E_7(X)\leqslant U$. Classify each of its 21 interaction pairs as \textit{strong} when $V(r)<-1/4$ and \textit{weak} otherwise, and let $w$ be the number of weak pairs. Since $V(r)=(r^{-6}-1)^2-1\geqslant-1$, each strong pair contributes at least $-1$ and each weak pair contributes at least $-1/4$. Hence,
\begin{equation}
E_7(X)\geqslant -1\cdot(21-w)-\tfrac{1}{4}w
=-21+\tfrac{3}{4}w.
\label{eq:weak}
\end{equation}
If $w\geqslant6$, then $E_7(X)\geqslant-16.5>U$. Thus every competitive configuration has at most five weak pairs. 

Let the \textit{strong graph} have the seven particles as vertices and the strong pairs as edges. Solving $V(r)<-1/4$ shows that every strong separation satisfies $a_+<r<a_-$, where $a_\pm=(1\pm\sqrt 3 /2)^{-1/6}$. Any two vertices are adjacent or have a common neighbour, otherwise their own pair and at least one pair to each of the other five vertices would be weak, giving at least six weak pairs. By the triangle inequality, every pair separation satisfies $r_{ij}<2a_-<2.8$. 

The repulsive core also gives a lower separation bound. Suppose some $r_{ij}\leqslant1/2$, then $V(r_{ij})\geqslant V(1/2)=3968$ while each of the 20 remaining pairs contributes at least -1, which gives $E_7(X)\geqslant3948$. Therefore, each competitive configuration must satisfy $1/2<r_{ij}<2.8$. 

The candidate $Y$ shows that the competitive set is nonempty. After fixing one particle at the origin to remove translations, place the configurations in the outer set $1/2\leqslant r_{ij}\leqslant2.8$, which is compact: the upper bounds make it bounded and the lower bounds make it closed and collision-free. The competitive set is the inverse image of $(-\infty,U]$ under the continuous energy on this outer set, hence is closed there and compact; the preceding strict estimates show that it meets neither boundary. Thus $E_7$ attains a minimum on the competitive set. Since $E_7(Y)<U$, any configuration with lower energy would itself be competitive. Therefore this minimum is the global minimum of $E_7$. Moreover, the minimiser is collision-free and lies in the unconstrained configuration space of distinct particles, and so satisfies $\nabla E_7(X)=0$.

Next we use the lower bounds for the smaller $N=5,6$ LJ clusters, i.e., 
\begin{equation}
    E_5\geqslant L_5 =-9.134,\quad E_6\geqslant L_6 = -13.2765.
\end{equation}
These bounds hold for any arrangement of five or six particles in $\mathbb{R}^3$. Thus, after deleting two particles from $X$, the five remaining particles have mutual energy at least $L_5$. Similarly, deleting one particle leaves a six-particle configuration with mutual energy at least $L_6$. We assume no stationarity or optimality of the remaining particles. The full proof is given in SM Sec.~S2. We use these bounds below to rule out several possible configurations with at most five weak pairs. 

We now encode the remaining combinatorial freedom by the weak interactions. Define the \textit{weak graph} $G_w(X)$ on the seven particles by joining $i$ and $j$ exactly when the pair $ij$ is weak. Since $w\leqslant5$, $G_w(X)$ has at most five edges (up to particle relabelling). Contact-graph enumeration has previously been used to classify finite rigid and sticky-sphere packings \cite{Arkus2011,HolmesCerfon2016}, while Gram-matrix formulations are standard in Euclidean distance geometry \cite{Liberti2014}. Here $G_w$ is not a contact model: it records an energy threshold only, and all 21 pair interactions remain in $E_7$. 

Up to graph isomorphism, there are exactly $40$ weak graphs on seven vertices with at most 5 edges. Specifically, there is 1 graph with 0 weak edges, 1 with 1 weak edge, 2 with 2, 5 with 3, 10 with 4 and 21 with 5 weak edges. The exhaustive count from Burnside's lemma and the complete set of graph representatives are given in SM Sec.~S3. 

We show that subset bounds eliminate 11 of these 40 possibilities immediately. First, if some particle belongs to at least four weak pairs, deleting that particle would leave a six-particle configuration with energy at least $L_6$. Of the 6 interactions involving the deleted particle, at least 4 are weak, so their total contribution is bounded below by $-2-4/4=-3$ and hence $E_7(X)\geqslant L_6-3=-16.2765>U$. Such a configuration cannot be competitive, and exactly 5 of the 40 graph types have a vertex of weak degree at least 4. 

Now consider a five-edge weak graph for which two vertices together meet every weak edge. Deleting these two particles leaves five particles containing no weak pairs, whose energy is at least $L_5$. Of the remaining 11 interactions, 5 are weak and 6 are strong, and therefore $E_7(X)\geqslant L_5-6-5/4=-16.384>U$. Thus such weak-pair arrangements are also impossible. This removes six additional graph types, so 11 of the 40 weak graphs are excluded, leaving 29 possibilities. 

\textit{Excluding short weak pairs. }The weak graph shows which pairs have energy at least $-1/4$, but it does not determine their range. Due to the repulsive core of the LJ potential, the same energy condition can occur at either a short or a long separation, i.e., 
\begin{equation}
    V(r)\geqslant-\tfrac14
\quad\Leftrightarrow\quad
r\leqslant a_+\ \text{ or }\ r\geqslant a_-,
\end{equation}
where $a_+\approx0.901$ and $a_-<1.398$. Hence, each weak pair could belong to either the short-distance repulsive branch or the long-distance attractive tail. Let $r_{\min}$ be the smallest pair separation. The force estimate of SM Sec.~S2 proves that every stationary configuration satisfies
\[
 r_{\min}\leqslant\frac{113}{125}
 \quad\Longrightarrow\quad
 E_7(X)>-16.49>U.
\]
This implication uses stationarity, already established for every global minimiser, but not for every competitive point. Therefore every global minimiser has $r_{\min}>113/125=0.904>a_+$, and each of its weak pairs lies on the long-distance branch $r_{ij}\geqslant a_-$. 

Thus each of the 29 remaining weak graphs specifies bounds on every pair separation. Together, these 29 bounded distance domains contain every global minimiser, up to particle relabelling.

\textit{Step 2: exclude the remaining domains and localise the competitor in the surviving domain. }Step 1 reduces every possible global minimiser, up to particle relabelling, to one of 29 bounded distance domains. These pairwise bounds are necessary but not sufficient as 21 numbers lying in the allowed intervals need not be the mutual distances of seven points in $\mathbb{R}^3$. We therefore need to add the necessary three-dimensional compatibility conditions. We formulate them as a linear relaxation, allowing each domain to be excluded by a short exact rational certificate and the exclusions then combined, rather than relying on a single optimum of a nonlinear numerical calculation. 

For each particle pair $e,f$, we let $v_e$ be its pair vector and set $s_e=|v_e|^2$. The matrix $v_e v_e^{\mathsf T}$ gives both the length and direction of the pair. If we remove the direction-independent part, we have the traceless tensor $T_e=v_ev_e^{\mathsf T}-s_eI_3/3$. The Gram matrix of these tensors must be positive semidefinite, i.e.
\begin{equation}
 K_{ef}=\operatorname{tr}(T_eT_f)
       =(v_e\cdot v_f)^2-\tfrac13s_es_f,
 \qquad K\succeq0.
 \label{eq:tensor}
\end{equation}
This condition captures some of the angular compatibility that is missing from the individual distance bounds, where the factor $1/3$ is specific to $\mathbb{R}^3$. 

To express these constraints linearly, we introduce 
\begin{equation}
   W_{ef}=s_es_f,\quad
\epsilon_e=\phi(s_e),\quad
\phi(s)=s^{-6}-2s^{-3}, 
\end{equation}
where $e$ and $f$ label the $21$ particle pairs. At a physical configuration these hold exactly, and $E_7=\sum_e\epsilon_e$. The pair-vector inner products are linear combinations of squared distances, so their squares become linear expressions in $W$. Consequently, for any fixed rational vector $\alpha$, the necessary condition $\alpha^{\mathsf T}K\alpha\geqslant0$ is a linear inequality in $(s,W)$.

The geometric constraints, $E_5$ and $E_6$ energy bounds, and bounds on the LJ potential can then all be expressed in terms of $z=(s,W,\epsilon)$. On each distance domain, we replace the nonlinear relations $W_{ef}=s_es_f$ and $\epsilon_e=\phi(s_e)$ by linear inequalities. The resulting linear problem contains every physical configuration, together with some additional nonphysical points. Therefore, if even this larger set has energy above $U$, then the original physical domain is impossible. 

We now only need to certify a lower bound for this enlarged linear problem. Write the inequalities as $\mathsf Mz\leqslant b, \ell\leqslant z\leqslant u,$ where $\ell,u$ are the variable bounds, and let $c$ be chosen such that 
\begin{equation}
    c^{\mathsf T}z=\sum_e\epsilon_e=E_7.
\end{equation}
Any nonnegative weighted sum of the valid inequalities is itself valid. Let $\mu\geqslant0$ denote these weights and define $\xi=c+\mathsf M^{\mathsf T}\mu$. Using also the bounds $\ell\leqslant z\leqslant u$, every feasible point satisfies
\begin{equation}
c^{\mathsf T}z\geqslant
-\mu^{\mathsf T}b+
\sum_i\min(\xi_i\ell_i,\xi_i u_i).
\label{eq:dual}
\end{equation}
Equation~\eqref{eq:dual} is the certificate here: if its right-hand side exceeds $U$, then even the enlarged linear problem has energy above $U$, and therefore no physical configuration in that domain can be competitive. Floating-point optimisation is used only to find the useful weights $\mu$. They are then replaced by exact rational values and~\eqref{eq:dual} is evaluated with exact arithmetic, with no floating-point rounding in the final bound. Thus, the numerical optimisation acts as a guide to the search process but is not itself part of the proof.

\textit{Reduction from 29 to 7. } We first test the $29$ graph domains without subdivision. The all-strong case can be excluded analytically. Writing $S_p=\sum_{i<j}s_{ij}^p$, the squared distances of seven points in $\mathbb{R}^3$ satisfy the distance-moment inequality
\begin{equation}
        S_2\geqslant\tfrac{8}{147}S_1^2.
\end{equation}
Combining with a quadratic lower bound on the LJ potential over the strong interval, this gives $E_7\geq -16.1>U$ (SM Sec.~S3).  

For $19$ further graph domains, we find rational weights $\mu$ for which the exact bound in Eq.~\eqref{eq:dual} exceeds $U$, excluding the entire domain. The graph-by-graph certificates and certified bounds are listed in SM Sec.~S3.

Two additional domains are closed by a stronger five-particle bound. In each case, deleting two particles leaves five particles whose ten mutual pairs are all strong. Since Step~1 also gives $r_{ij}>0.904$, SM Sec.~S3 proves that this five-particle subset has energy greater than \(-8.5\). Among the other $11$ interactions, four are weak and seven are strong, so $E_7>-8.5-7-\frac44=-16.5>U.$ These three mechanisms exclude $22$ of the $29$ graph domains, leaving seven. 

\textit{Reduction from 7 to 1. } For the remaining seven domains, we subdivide selected distance intervals and apply the same exact certification to every resulting subdomain. We exploit relabellings that preserve the weak graph and the current distance bounds, and hence subdomains related by such a relabelling are equivalent and need be checked only once. This bookkeeping makes no symmetry assumptions on the particles. The subdivision excludes six further graph types, with every certified lower bound lying above \(U\). The smallest exclusion bound is $-16.504061292892>U.$

Only one weak graph survives, $C_5\sqcup2K_1,$ consisting of a five-cycle and two isolated vertices (SM Sec.~S3). In the PBP, the five weak pairs are the ring diagonals and the isolated vertices are the two poles. The graph alone, however, does not imply a regular planar ring or symmetric pole positions.

We therefore use the surviving domain to bound how far a competitive configuration can lie from $Y$. Let \(\bar s_e\) be rational approximations to the \(21\) squared pair distances of \(Y\), with their certified approximation error given in SM Sec.~S4, and define
\begin{equation}
D_{\rm ref}=\sum_{e=1}^{21}(s_e-\bar s_e)^2 .
\label{eq:deviation}
\end{equation}
We now add the competitive-energy condition \(E_7\leqslant U\) and apply the same exact certificate framework to the surviving domain. Successive certificates tighten the allowed distance intervals and ultimately give
\begin{equation}
D_{\rm ref}\leqslant 0.000022533598161300.
\label{eq:localization}
\end{equation}
Thus every global minimiser has its full set of \(21\) squared pair distances confined to a small neighbourhood of those of \(Y\), without assuming planarity, equal ring distances, or PBP symmetry.

\emph{Step 3: turning proximity into exact uniqueness. }
Step~2 controls the pair distances of every possible global minimiser,
whereas the final rigidity argument requires a bound on its Cartesian
displacement from \(Y\). Since pair distances are unchanged by
translations, rotations and reflections, we first centre both
configurations to remove translations.

For a centred configuration, the squared pair distances determine its
Gram matrix \(XX^{\mathsf T}\). Thus control of all 21 squared distances
can be converted into control of the corresponding Gram matrix, which
describes the geometry independently of its orientation.

Step~2 localises the squared distances relative to the rational reference
values \(\bar s_e\), whereas the rigidity argument must compare \(X\)
with the exact PBP \(Y\). Let \(s_e^\ast\) denote the exact squared pair
distances of \(Y\). The reference values satisfy the certified error bound
\[
 \sum_e(\bar s_e-s_e^\ast)^2<2\times10^{-24}.
\]
Combining this reference error with Eq.~\eqref{eq:localization}, and using
the linear relation between squared-distance matrices and centred Gram
matrices, gives
\begin{equation}
 \left\|XX^{\mathsf T}-YY^{\mathsf T}\right\|_F
 <\varepsilon=\frac{17}{5000}.
 \label{eq:gramerror}
\end{equation}
Here \(\|\cdot\|_F\) denotes the Frobenius norm. The derivation is given
in SM Sec.~S4.

The Gram matrix is invariant under \(X\mapsto XO\), where
\(O\in O(3)\) is a rotation or reflection. We therefore compare \(X\)
with \(Y\) after choosing the orthogonal transformation that aligns them
most closely. Since the PBP spans three-dimensional space, \(Y\) has
full column rank. More generally, if
\(\lambda_{\min}(Y^{\mathsf T}Y)\geqslant\lambda_0>\varepsilon\), then
the Gram-matrix bound implies
\begin{equation}
 \Lambda:=\min_{O\in O(3)}\|XO-Y\|_F^2
 \leqslant
 \frac{\varepsilon^2}{2(\lambda_0-\varepsilon)}.
 \label{eq:alignmentlemma}
\end{equation}
This alignment lemma is proved in SM Sec.~S4. For the PBP, we certify
\[
 \lambda_{\min}(Y^{\mathsf T}Y)>\lambda_0=\frac{13}{25}.
\]
Together with \(\varepsilon=17/5000\), this gives
\[
 \Lambda\leqslant\frac{289}{25830000}
 <R^2,\qquad R=\frac1{250}.
\]
Hence every global minimiser can be rotated or reflected so that its
Cartesian distance from \(Y\) is less than \(R\).

\textit{Local rigidity. }
Proximity is not yet equality. We therefore need a positive
restoring-energy bound throughout this ball, not merely positive
curvature at its centre.

Collect the Cartesian coordinates of the seven particles into a vector
in \(\mathbf R^{21}\), and let \(H(Z)\) denote the \(21\times21\)
Hessian of \(E_7\) at a configuration \(Z\). Since the energy is
invariant under translations and rotations, there are six infinitesimal
rigid-motion directions: three translations and three rotations. Let
\(T\) be the \(21\times6\) matrix whose columns span these directions
at \(Y\), and keep \(T\) fixed throughout the argument. A displacement
\(v\) satisfying \(T^{\mathsf T}v=0\) is therefore orthogonal to every
infinitesimal rigid motion at \(Y\).

The interval estimates of SM Sec.~S5, together with the comparison in
End Matter, give the uniform bound
\begin{equation}
 \begin{gathered}
 v^{\mathsf T}H(Z)v\geqslant 11.80808\,\|v\|^2,\\
 \|Z-Y\|_F\leqslant R,\qquad T^{\mathsf T}v=0.
 \end{gathered}
 \label{eq:uniform}
\end{equation}
Thus the energy has uniformly positive curvature throughout the ball
on the fixed slice orthogonal to the rigid motions at \(Y\).

Choose \(O_\star\in O(3)\) attaining the minimum in the definition of
\(\Lambda\), and set
\[
 \delta=XO_\star-Y.
\]
Then \(\|\delta\|_F<R\). Since \(X\) and \(Y\) are centred, \(\delta\)
is orthogonal to the three translation directions. Moreover,
first-order optimality of \(O_\star\) implies that
\(\delta^{\mathsf T}Y\) is symmetric, which is equivalent to
orthogonality to the three infinitesimal rotation directions at \(Y\).
Hence
\[
 T^{\mathsf T}\delta=0.
\]

The straight path \(Z(t)=Y+t\delta\), \(0\leqslant t\leqslant1\),
remains inside the ball because
\[
 \|Z(t)-Y\|_F=t\|\delta\|_F<R.
\]
Since \(T\) is fixed at \(Y\), the condition
\(T^{\mathsf T}\delta=0\) holds along the whole path, so
Eq.~\eqref{eq:uniform} applies to the same displacement \(\delta\) at
every point.

Finally, \(Y\) is stationary, so \(\nabla E_7(Y)=0\). Taylor's formula
therefore gives
\begin{equation}
 \begin{aligned}
 E_7(Y+\delta)-E_7(Y)
 &=\int_0^1(1-t)\,
   \delta^{\mathsf T}H(Y+t\delta)\delta\,dt\\
 &\geqslant\frac{59}{10}\|\delta\|_F^2.
 \end{aligned}
 \label{eq:rigidity}
\end{equation}
Hence every nonzero aligned displacement from \(Y\) in this
rigid-motion-free slice has strictly higher energy.

Every global minimiser can be aligned into this neighbourhood by the
preceding localisation and alignment argument, and that alignment
leaves its energy unchanged. Since a global minimiser cannot have
energy greater than \(E_7(Y)\), Eq.~\eqref{eq:rigidity} forces
\(\delta=0\). Therefore every global minimiser is congruent to \(Y\),
up to particle relabelling. This proves both global optimality and
uniqueness.

\textit{The $N=7$ Thomson problem. }The spherical PBP is a classical candidate, appearing in early stability studies and later numerical catalogues of equilibrium and apparent global-minimum configurations \cite{Foppl1912,Cohn1956,Melnyk1977,ErberHockney1991,Altschuler1997}. More recently, Kuznetsov and Sahinidis certified global optimality, within numerical tolerance, of the putative seven-electron solution \cite{KuznetsovSahinidisThomson2026}. Here we strengthen that numerical result to exact global optimality and uniqueness. For seven unit vectors $X=(x_0,\ldots,x_6)\in(S^2)^7$, define the
Thomson energy
\begin{equation}
 \mathcal E_{\rm T}(X)=\sum_{i<j}\frac1{\norm{x_i-x_j}}.
 \label{eq:t-energy}
\end{equation}
Let $Y_{\rm T}=(y_0,\ldots,y_6)$ be the unit-sphere PBP, with five equally spaced equatorial
vertices and two poles, and set
\begin{equation}
 D_{\rm T}(X)=\sum_{i<j}
 \left(\norm{x_i-x_j}^2-\norm{y_i-y_j}^2\right)^2.
 \label{eq:t-D}
\end{equation}
The exact candidate energy is
\begin{equation}
 \mathcal E_{\rm T}(Y_{\rm T})=
 \tfrac12+5\sqrt2+5\sqrt{1+2/\sqrt5}.
\end{equation}
Rigorous finite-$N$ spherical-energy methods include auxiliary-function bounds, universal optimality, localisation, and higher-point semidefinite bounds \cite{Yudin1993,CohnKumar2007,Schwartz2013,CohnWoo2012}. Very recently, the eight-point Coulomb and logarithmic problems were settled by a fully Lean-verified computer-assisted proof \cite{KryvonosLiehrTaylor2026}.

The proof is expressed through two theorems.
\emph{Global-entry theorem.} Every collision-free spherical
stationary configuration satisfying
$\mathcal E_{\rm T}(X)\leqslant\mathcal E_{\rm T}(Y_{\rm T})$ can be
relabelled into one explicit PBP-compatible domain $\mathcal K_{\rm T}$.
This single theorem consolidates the harmonic, separation, three-point,
long-edge-graph, pole, and ring-height reductions; its three-part proof is given
in SM Sec.~S8.

\emph{Whole-domain coercivity theorem.} Every physical configuration in
$\mathcal K_{\rm T}$ satisfies the whole-domain geometric inequality, where for $ X\in\mathcal K_{\rm T}$, we have
\begin{equation}
 \mathcal E_{\rm T}(X)-\mathcal E_{\rm T}(Y_{\rm T})
 \geqslant \tfrac{1}{4000}D_{\rm T}(X).
 \label{eq:t-coercivity}
\end{equation}
It assumes neither stationarity nor symmetry and includes $D_{\rm T}=0$.
The proof uses candidate-sharp quartic Coulomb minorants and common
rank-three Gram geometry; End Matter gives the short leakage--stress
completion, while the Supplemental Material records all exact factor and
verification interfaces.

A global minimiser exists, is collision-free and stationary, and has energy
at most that of $Y_{\rm T}$.  The global entry theorem therefore puts it in
$\mathcal K_{\rm T}$, after which Eq.~\eqref{eq:t-coercivity} gives
\[
 0\geqslant \mathcal E_{\rm T}(X)-\mathcal E_{\rm T}(Y_{\rm T})
 \geqslant  \tfrac{1}{4000}D_{\rm T}(X)\geqslant0.
\]
Thus $D_{\rm T}=0$, the full uncentred Gram matrix agrees with that of the
PBP, and $X$ is congruent to $Y_{\rm T}$ up to $O(3)$ and particle
relabelling.  The domain-coercivity and harmonic checks are freshly
replayable; the finite global-entry archives retain the verification boundary
stated in the verification section of the Supplemental Material.


The LJ theorem gives a complete zero-temperature classification of the
seven-particle LJ cluster, i.e., an isotropic pair interaction selects the
fivefold PBP even when every distorted and asymmetric competitor is
allowed. This provides an exact finite-cluster example of structure
emerging from geometric frustration among all 21 pair interactions,
rather than from an imposed symmetry or a directional interaction.

More broadly, the proof provides a modular route from numerical
structure prediction to rigorous finite-\(N\) classification. Coarse
energy bounds first reduce the continuous configuration space to
finitely many interaction topologies; three-dimensional distance
geometry and exact rational certificates then exclude incompatible
domains; and localisation together with uniform stiffness closes the
surviving equality case. The separation between geometric constraints
and potential-specific energy bounds suggests how the same architecture
could be adapted to other particle numbers and other isotropic pair
potentials. Such extensions are not automatic, since the number of
competing domains grows rapidly and stronger global reductions will be
needed. Nevertheless, the LJ$_7$ result demonstrates a concrete
framework for converting a numerical cluster candidate into an exact,
symmetry-unrestricted ground-state classification. The Thomson companion
shows that the final localisation-and-stiffness step can sometimes be
replaced by one global-entry theorem followed by a rank-three coercivity
identity.

\paragraph*{Use of generative artificial intelligence.}
Generative AI tools (GPT 5.6 and 6 Astra) were used to assist with mathematical exploration, code development, debugging and verification workflows. All final mathematical claims, proofs, computations, code, and certificates were reviewed and validated by the author, who takes responsibility for the content.

\label{main:last}
\end{document}